\documentclass[prd,twocolumn,showpacs,preprintnumbers,amsmath,amssymb,superscriptaddress,floatfix,nofootinbib]{revtex4}

\usepackage{graphicx}
\usepackage{amsmath}
\usepackage{amsfonts}
\usepackage{amssymb}
\usepackage{color}
\usepackage{multirow}
\usepackage[colorlinks, citecolor=blue,anchorcolor=red,menucolor=red,linkcolor=red,filecolor=red,runcolor=red,urlcolor=blue,frenchlinks=red]{hyperref}

\usepackage{subfigure}

\begin{document}

\title{Search for a $D \bar{D}$ bound state in the $\Lambda_b \rightarrow \Lambda D\bar{D}$ process}

\author{Le-Le Wei}
\affiliation{School of Physics and Microelectronics, Zhengzhou University, Zhengzhou, Henan 450001, China}

\author{Hong-Shen Li}
\affiliation{School of Physics and Microelectronics, Zhengzhou University, Zhengzhou, Henan 450001, China}

\author{En Wang}
\email{wangen@zzu.edu.cn}
\affiliation{School of Physics and Microelectronics, Zhengzhou University, Zhengzhou, Henan 450001, China}

\author{Ju-Jun Xie}\email{xiejujun@impcas.ac.cn}
\affiliation{Institute of Modern Physics, Chinese Academy of
Sciences, Lanzhou 730000, China} \affiliation{School of Nuclear
Sciences and Technology, University of Chinese Academy of Sciences,
Beijing 101408, China} \affiliation{School of Physics and
Microelectronics, Zhengzhou University, Zhengzhou, Henan 450001,
China}

\author{De-Min Li}
\affiliation{School of Physics and Microelectronics, Zhengzhou University, Zhengzhou, Henan 450001, China}

\author{Yu-Xiao Li}
\affiliation{School of Physics and Microelectronics, Zhengzhou University, Zhengzhou, Henan 450001, China}

\begin{abstract}

We have investigated the process of $\Lambda_b\to \Lambda D\bar{D}$, by taking into account the contributions from the $s$-wave $D\bar{D}$ interaction within the coupled-channel unitary approach, and the intermediate $\psi(3770)$ resonance. In addition to the peak of the $\psi(3770)$, an enhancement near the $D\bar{D}$ mass threshold is found in the $D\bar{D}$ invariant mass distributions, which should be the reflection of the $D\bar{D}$ bound state. We would like to encourage our experimental colleagues to measure the $D\bar{D}$ invariant mass distribution of the $\Lambda_b\to \Lambda D\bar{D}$ process, which is crucial to search for the $D\bar{D}$ bound state and to understand the heavy-hadron heavy-hadron interactions.
\end{abstract}

%\pacs{Valid PACS appear here}
% PACS, the Physics and Astronomy Classification Scheme.
% Valid PACS numbers may be entered using the \verb+\pacs{#1} command.

%\keywords{Baryons, Mesons, Resonances, Bound states, Chiral unitary approach, Nonperturbative technique.}

\maketitle

%%%%%%%%%%%%%%%%%%%%%%
\section{Introduction} \label{sec:Introduction}

Although the quark model was proposed by Gell-Mann and Zweig more than half century ago~\cite{GellMann:1964nj,Zweig:1964jf}, it is still valid in classifying all known hadrons by now. Since the $X(3872)$ was observed by the Belle Collaboration in 2003~\cite{Choi:2003ue}, many charmonium-like states were reported experimentally~\cite{Zyla:2020zbs}, and most of them cannot be explained as the conventional mesons ($q\bar{q}$) or baryons ($qqq$)~~\cite{Olsen:2017bmm,Brambilla:2019esw}.
There are many explanations about those states, such as tetraquark states, molecular states, the conventional $c\bar{c}$ mesons, or the mixing between different components~\cite{Chen:2016qju,Liu:2019zoy,Hosaka:2016pey,Guo:2017jvc,Hao:2019fjg}.  However, it is surprising that many resonant structures are observed around thresholds of a pair of heavy hadrons, such as $X(3872)$ and $Z_c(3900)^\pm$ around the $D\bar{D}^*$ threshold, $Z_{cs}(3985)$ around the $\bar{D}_sD^*$ and $\bar{D}^*_s D$ thresholds, and $X(3930)$ around $D_s\bar{D}_s$ threshold. As discussed in Ref.~\cite{Dong:2020hxe}, such  structures should appear at any  threshold of a pair of heavy-quark and heavy-antiquark hadrons which have attractive interaction at threshold. Thus, the experimental information about the threshold structures is crucial to deeply understand the heavy-hadron heavy-hadron interactions, and the internal structures of the hidden-charm states~\cite{Wang:2018djr,Wang:2017mrt}.

In Ref.~\cite{Gamermann:2006nm}, one new hidden charm resonance with mass around 3700~MeV (denoted as $X(3700)$ in this article) is predicted within the coupled channel unitary approach involving the $D^+D^-$, $D^0\bar{D}^0$, $D_s\bar{D}_s$, $K^+K^-$, $K^0\bar{K}^0$, $\pi^+\pi^-$, $\pi^0\pi^0$, $\eta\eta$, and $\pi^0\eta$ channels. Later it was suggested to search for this predicted $D\bar{D}$ bound state in several processes, such as  $B\to D\bar{D}K$~\cite{Dai:2015bcc}, $\psi(3770)\to \gamma X(3700) \to \gamma \eta \eta'$, $\psi(4040)\to \gamma X(3700) \to \gamma \eta\eta'$, and $e^+e^-\to J/\psi X(3700)\to J/\psi  \eta\eta'$~\cite{Xiao:2012iq}. According to the studies of Refs.~\cite{Gamermann:2007mu,Wang:2019evy}, the experimental data of $e^+e^-\to J/\psi D\bar{D}$ measured by the Belle Collaboration~\cite{Abe:2007sya,Chilikin:2017evr} are compatible with the existence of such a $D\bar{D}$ bound state around 3700~MeV, though other possibilities cannot be discarded due to the present quality of the Belle data. In Ref.~\cite{Wang:2020elp}, we have performed a global fit to the data of $\gamma\gamma \to D\bar{D}$~\cite{Uehara:2005qd,Aubert:2010ab} and the $e^+e^-\to  J/\psi D\bar{D}$~\cite{Chilikin:2017evr}, by taking into account the $s$-wave $D\bar{D}$ final state interactions. Our results are consistent with the experimental data considering the uncertainties of the fitted parameters,  and the modulus squared of the amplitude $|t_{D\bar{D}\to D\bar{D}}|^2$ show peaks around $3710\sim 3740$~MeV~\cite{Wang:2020elp}. Recently, a $D\bar{D}$ bound state with binding energy $B=4.0^{+5.0}_{-3.7}$~MeV was also predicted according to the Lattice calculation in Ref.~\cite{Prelovsek:2020eiw}. Thus, it is crucial to search for the signal of this predicted state.

On the other hand, the decays of $\Lambda_b$ is one of the important tool to study the hidden charm resonances~\cite{Oset:2016lyh}, such as the processes of $\Lambda_b \to J/\psi \Lambda$, $\Lambda_b\to \psi(2S)\Lambda$~\cite{Aaij:2019dvk,Abazov:2011wt,Aad:2015msa}. The process $\Lambda_b \to \Lambda  X^0_c$ ($X^0_c \equiv c\bar{c}u\bar{u}(d\bar{d)}, c\bar{c}s\bar{s}$) is also proposed to search for the $XYZ$ states in Ref.~\cite{Hsiao:2015txa}.  In this work, we will propose to search for the signal of the $D\bar{D}$ bound state in the single-Cabibbo-suppressed process of  $\Lambda_b\to \Lambda D\bar{D}$, which has not been measured experimentally up to our knowledge. It should be pointed out that the $\Lambda_b\to \Lambda D\bar{D}$ process is expected to  have a larger branching fraction than the double-Cabibbo-Suppressed process $\Lambda_b\to \Lambda K^+K^-$ with the branching fraction $\mathcal{B}(\Lambda_b\to \Lambda K^+K^-)=(15.9\pm1.2\pm1.2\pm2.0)\times 10^{-6}$ measured by the LHCb Collaboration~\cite{Aaij:2016nrq}.

Since the predicted mass of the $D\bar{D}$ bound state is lower than the $D\bar{D}$ threshold,
it will manifest itself as the enhancement near the $D\bar{D}$ threshold, the similar work is found in Refs.~\cite{Dai:2015bcc,Wang:2020wap}.
For instance, a peak observed in the $\phi\omega$ threshold in the $J/\psi\to \gamma \phi\omega$ reaction~\cite{Ablikim:2006dw} was interpreted as the manifestation of the $f_0(1710)$ resonance below the $\phi\omega$ threshold~\cite{Geng:2008gx}. In Ref.~\cite{Ablikim:2009ac} the BESIII Collaboration has seen a bump structure close to threshold in the $K^{*0}\bar{K}^{*0}$ mass distribution of the $J/\psi\to \eta K^{*0}\bar{K}^{*0}$ decay, which can be interpreted as a signal of the formation of an $h_1$ resonance~~\cite{Xie:2013ula,Geng:2008gx}.  We expect there will be an enhancement near the threshold in the $D\bar{D}$ invariant mass distribution. On the other hand, since the $\psi(3770)$, with a mass close to the $D\bar{D}$ threshold, mainly decays into $D\bar{D}$ in $p$-wave, we will take into account the contribution from the $\psi(3770)$.

The paper is organized as follows. In Sect.~\ref{sec:Formalism}, we introduce our model for  the process $\Lambda_b \rightarrow \Lambda D\bar{D}$. Numerical results for the $D \bar{D}$  invariant mass distribution and discussions are given in Sect.~\ref{sec:Results}, and a short summary is given in the last section.

%%%%%%%%%%%%%%%%%%%%%%
\section{Formalism}  \label{sec:Formalism}

\begin{figure}[tbhp]
\begin{center}
\includegraphics[scale=0.7]{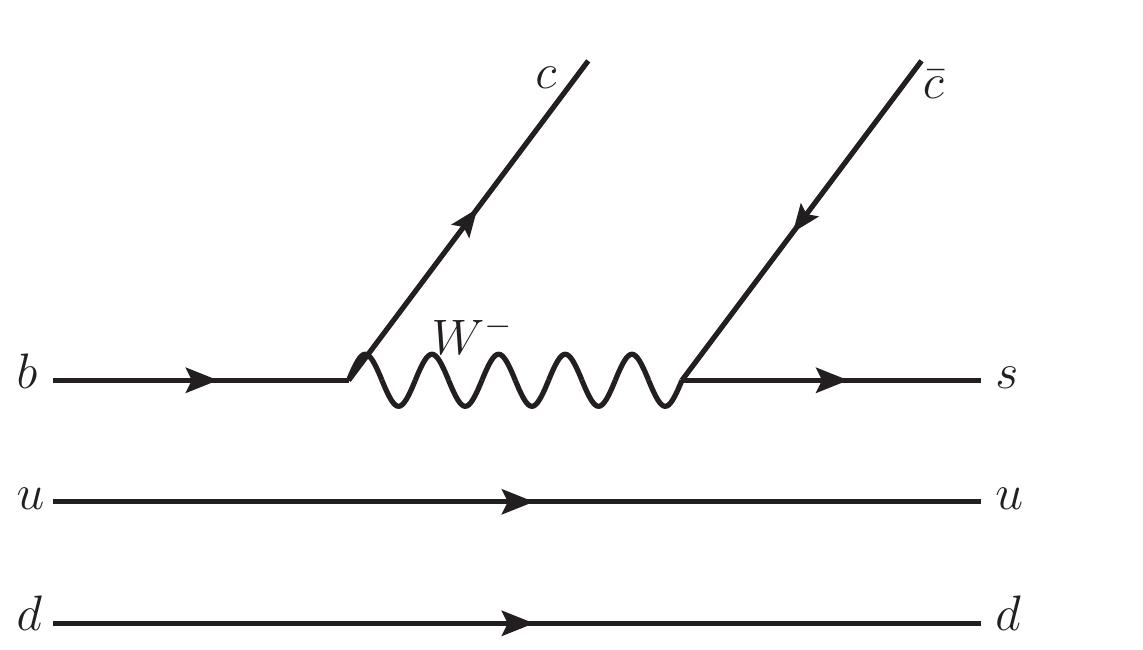}
\end{center}
%\vspace{-0.7cm}
\caption{The quark level diagram for the weak decay $\Lambda_b \rightarrow \Lambda c\bar{c}$.}
\label{Fig:QuarkLevel}
\end{figure}

In analogy to Refs.~\cite{Li:2020fqp, Wang:2020pem, Wang:2015pcn, Lu:2016roh, Liu:2020ajv}, the mechanism of the decay $\Lambda_b \rightarrow \Lambda D \bar{D}$ ($D\bar{D}\equiv D^0\bar{D}^0, D^{+} D^{-}$) can happen via three steps: the weak decay, hadronization, and the final state interaction. In the first step as depicted in Fig.~\ref{Fig:QuarkLevel}, the $b$ quark of the initial $\Lambda_b$ weakly decays into a $c$ quark and a $W^{-}$ boson, followed by the $W^{-}$ boson decaying into a $\bar{c} s$ quark pair,
\begin{eqnarray}
\left| \Lambda_b \right\rangle &=&  \frac{1}{\sqrt{2}} b(ud-du) \nonumber \\
&\Rightarrow& V_p c \bar{c} \frac{1}{\sqrt{2}} s (ud-du) \nonumber \\
&=& V_p c \bar{c} \Lambda ,
\label{Eq:HLambdab}
\end{eqnarray}
where we take the flavor wave functions $\Lambda_b = b\left(ud-du\right)/{\sqrt{2}}$ and $ \Lambda  =  s\left(ud-du\right)/{\sqrt{2}}$, and $V_p$ is the strength of the production vertex that contains all dynamical factors.

\begin{figure}[tbhp]
\begin{center}
\includegraphics[scale=0.7]{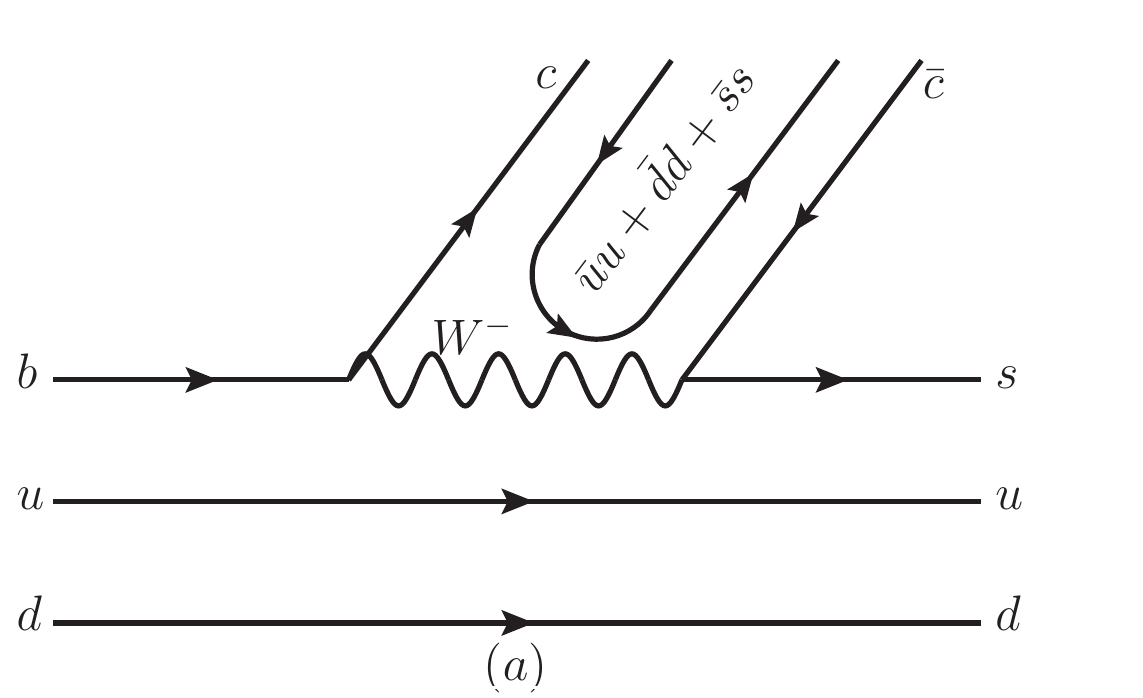}
\includegraphics[scale=0.7]{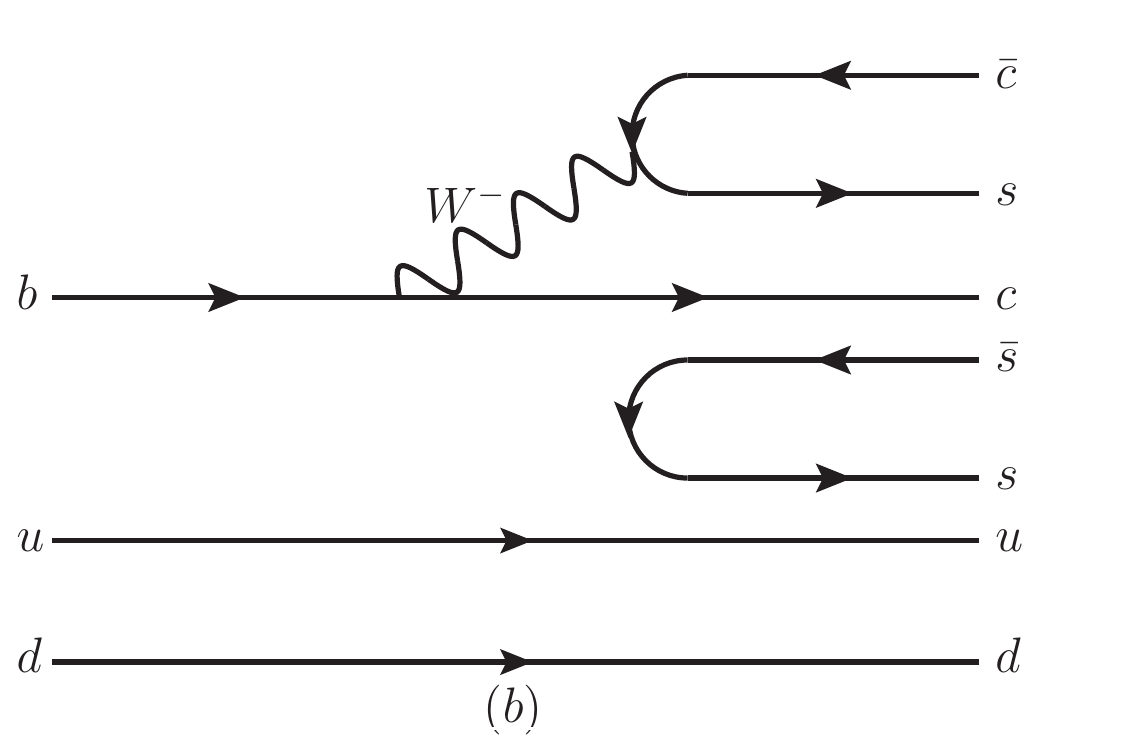}
\end{center}
%\vspace{-0.7cm}
\caption{The mechanisms of (a) the internal $W^-$ emission mechanism and (b) the external $W^-$ emission for the weak decay $\Lambda_b$ and the hadronization of the $c\bar{c}$ through $\bar{q}q$ created from the vacuum.} \label{Fig:hadronization}
\end{figure}

In order to give rise to the final state $D^0 \bar{D}^0 \Lambda$ (or $ D^+ D^- \Lambda$), the  quark $c$ and antiquark $\bar{c}$ need to hadronize together with the $\bar{q} q$ ($\equiv \bar{u} u + \bar{d} d + \bar{s} s$) created from the vacuum with $J^{PC}=0^{++}$, which could be expressed as  the mechanisms of the internal $W^-$ emission and external $W^-$ emission, respectively shown in Figs.~\ref{Fig:hadronization}(a) and \ref{Fig:hadronization}(b). Thus, we have,
\begin{eqnarray}
\left| H \right\rangle^{\rm in} &=& V_p \left| c \left( \bar{u} u + \bar{d} d + \bar{s} s \right) \bar{c} s\frac{1}{\sqrt{2}}\left(ud-du\right) \right\rangle \nonumber \\
&=& V_p \left( D^0 \bar{D}^0 + D^{+} D^{-} + D_s^{+} D_s^{-} \right) \Lambda ,
\label{Eq:Hinternal}
\end{eqnarray}
for the internal $W^-$ emission mechanism of Fig.~\ref{Fig:hadronization}(a) , and
\begin{eqnarray}
\left| H \right\rangle^{\rm ex} &=& V_p\times C\times D_s^{+} D_s^{-}  \Lambda ,
\label{Eq:Hinternal}
\end{eqnarray}
for the external $W^-$ emission mechanism of Fig.~\ref{Fig:hadronization}(b). Here the color factor $C$ accounts for the relative weight of the external $W^-$ emission with respect to the internal $W^-$ emission, and we take $C=3$ in the case of color number $N_c=3$~\cite{Duan:2020vye,Zhang:2020rqr,Dai:2018nmw}.

\begin{figure}[tbhp]
\begin{center}
\includegraphics[scale=0.7]{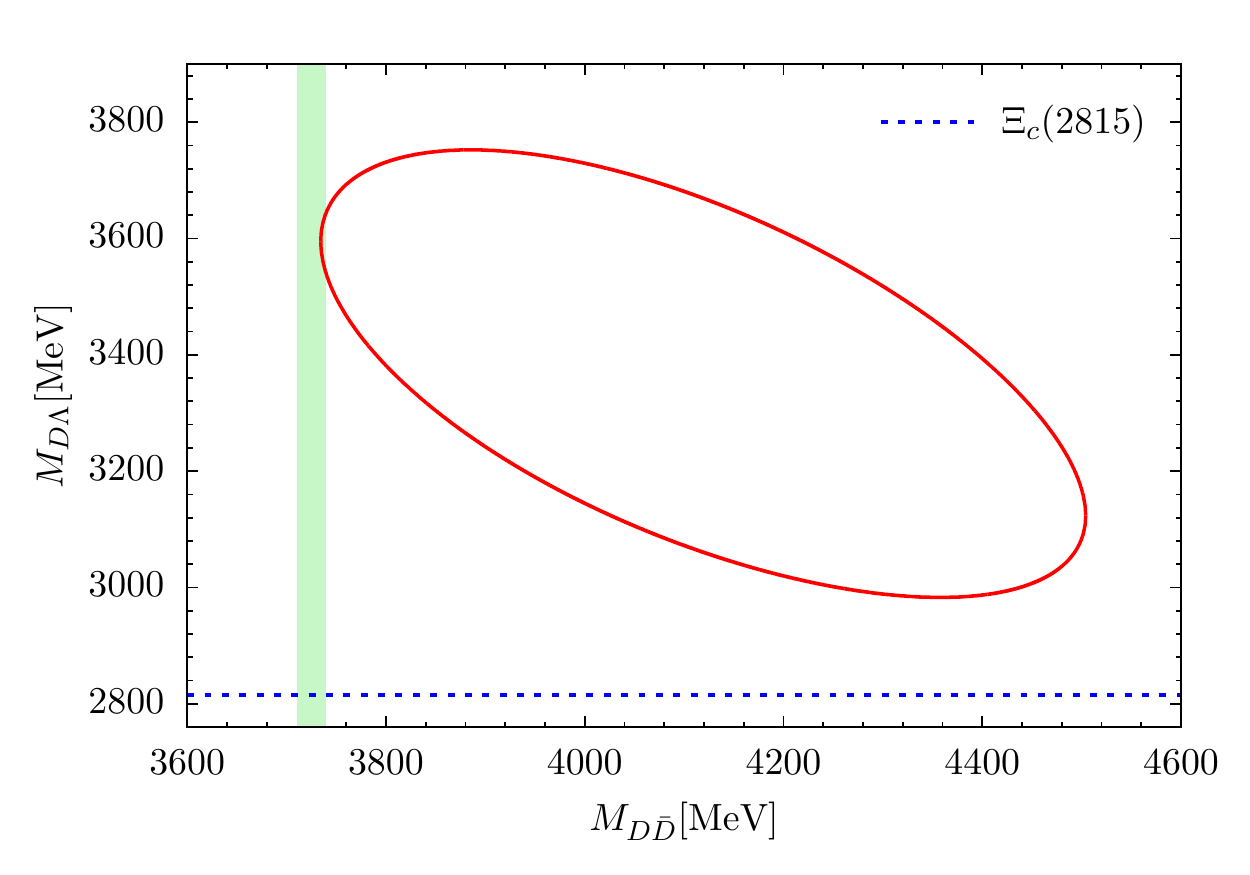}
\end{center}
%\vspace{-0.7cm}
\caption{The Dalitz plot for the $\Lambda_b \rightarrow  \Lambda D \bar{D}$. The green band stands for the region of $3710 \sim 3740$~MeV that the predicted $D\bar{D}$ bound state lies in.}
\label{Fig:Dalitz}
\end{figure}

The final states can also undergo the interactions of the $D\bar{D}$ and $\Lambda D$, which may generate dynamically the resonances. The interaction of the coupled channels including $\Lambda D $ was studied within a unitary coupled-channel approach which incorporates heavy-quark spin symmetry, and two resonances $\Xi_c(2790)$ and $\Xi_c(2815)$ are identified as the dynamically generated resonances~\cite{Romanets:2012hm}. Since their masses are about $150\sim 200$~MeV below the $\Lambda D$ threshold, their contributions do not affect the structure close to the $D\bar{D}$ threshold, which can be easily understood from the Dalitz plot of Fig.~\ref{Fig:Dalitz}. Thus, we neglect the $\Lambda D$ interaction in this work, because only the $D\bar{D}$ invariant mass distribution near the threshold is relevant for the $D\bar{D}$ bound state.

\begin{figure}[htbp]
\centering
  \includegraphics[scale=0.9]{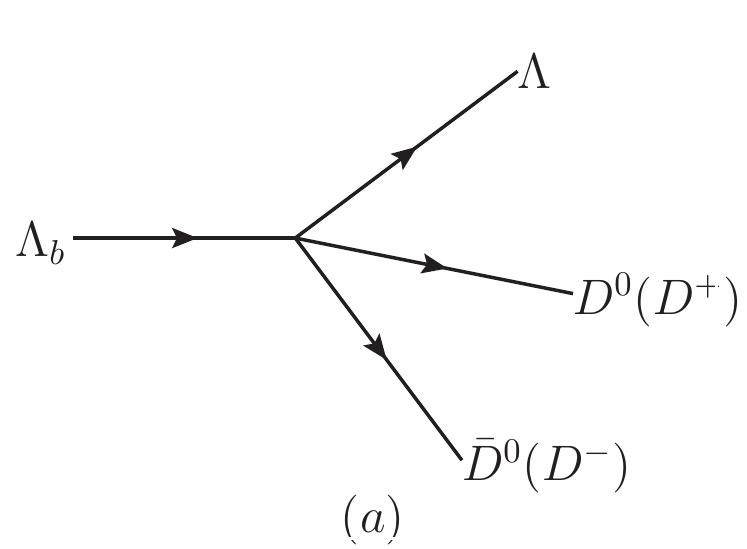} \\
  \includegraphics[scale=0.9]{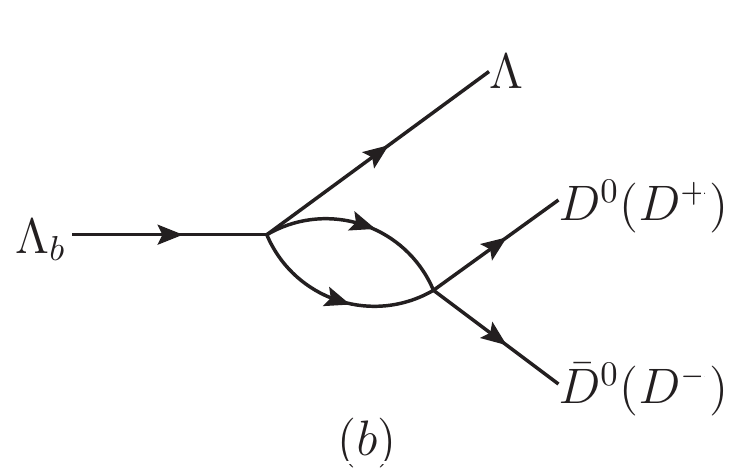}
\caption{The decays $\Lambda_b \rightarrow \Lambda D^0 \bar{D}^0$ and $\Lambda_b \rightarrow \Lambda D^{+} D^{-}$, (a) direct production, (b) the re-scattering of the channels  $D^0 \bar{D}^0$, $D^{+} D^{-}$, or $D_s^{+} D_s^{-}$.} \label{Fig:Feynman}
\end{figure}

The next step is to consider the final state interaction of these channels to give $D^0 \bar{D}^0$ (or $D^+ D^-$) at the end. We can have the final states of  $D^0 \bar{D}^0$ (or $D^+ D^-$) through the direct production in the $\Lambda_b$ decay, or  the re-scattering of the primarily produced channels $D^0 \bar{D}^0$, $D^{+} D^{-}$, or $D_s^{+} D_s^{-}$, as shown in Figs.~\ref{Fig:Feynman}(a) and \ref{Fig:Feynman}(b), respectively. Apart from the three coupled channels $D^0 \bar{D}^0$, $D^{+} D^{-}$, and $D_s^{+} D_s^{-}$, we only consider one light channel $\eta \eta$ to account for the width of the $D \bar{D}$ bound state, as in Refs.~\cite{Wang:2019evy, Xiao:2012iq, Dai:2015bcc, Wang:2020elp}.

Then, the total amplitudes for the $\Lambda_b \rightarrow \Lambda D^0 \bar{D}^0$ and $\Lambda_b \rightarrow \Lambda D^{+} D^{-}$ can be expressed as,
\begin{eqnarray}
t_{\Lambda_b \rightarrow \Lambda D^0 \bar{D}^0}^{s \text{-wave}} &=& V_p \left[ 1 + G_{D^{+} D^{-}} t_{D^{+} D^{-} \rightarrow D^0 \bar{D}^0} \right. \nonumber \\
&& + G_{D^0 \bar{D}^0} t_{D^0 \bar{D}^0 \rightarrow D^0 \bar{D}^0} \nonumber \\
&& \left. + (1+C) G_{D_s^{+} D_s^{-}} t_{D_s^{+} D_s^{-} \rightarrow D^0 \bar{D}^0} \right],
\label{Eq:tD0} \\
t_{\Lambda_b \rightarrow \Lambda D^{+} D^{-}}^{s \text{-wave}} &=& V_p \left[ 1 + G_{D^{+} D^{-}} t_{D^{+} D^{-} \rightarrow D^{+} D^{-}} \right. \nonumber \\
&& + G_{D^0 \bar{D}^0} t_{D^0 \bar{D}^0 \rightarrow D^{+} D^{-}} \nonumber \\
&&\left. + (1+C) G_{D_s^{+} D_s^{-}} t_{ D_s^{+} D_s^{-} \rightarrow D^{+} D^{-}} \right],
\label{Eq:tDp}
\end{eqnarray}
where $G_{l}$ is the loop function for the two-meson propagator in the $l$-th channel,
\begin{eqnarray}
G_{l} &=& i \int \frac{d^4 q}{(2\pi)^4} \frac{1}{q^2 - m_1^2 + i\epsilon} \frac{1}{(P-q)^2 - m_2^2 + i\epsilon} \nonumber \\
&=& \frac{1}{16\pi^2} \left[\alpha_l + \ln{\frac{m_1^2}{\mu^2}} + \frac{m_2^2 - m_1^2 + s}{2s} \ln{\frac{m_2^2}{m_1^2}} \right. \nonumber \\
&& + \frac{p}{\sqrt{s}} \times \left(\ln{\frac{s - m_2^2 + m_1^2 + 2p\sqrt{s}}{-s + m_2^2 - m_1^2 + 2p \sqrt{s}}} \right. \nonumber \\
&& \left. \left. + \ln{\frac{s + m_2^2 - m_1^2 + 2p\sqrt{s}}{-s - m_2^2 + m_1^2 + 2p \sqrt{s}}} \right) \right],
\label{Eq:LoopFuntion}
\end{eqnarray}
with the subtraction constant $\alpha_l= -1.3$ ($l = 1, 2, 3, 4$ correspond to the channels $D^0 \bar{D}^0, D^{+} D^{-}$, $D_s^{+} D_s^{-}$, and $\eta\eta$, respectively) and $\mu$ = 1500 MeV as  Ref.~\cite{Gamermann:2006nm}. $P\equiv \sqrt{s}=  M_{D\bar{D}}$ is the invariant mass of the two mesons in the $l$-th channel. $m_1$ and $m_2$ are the masses of the two mesons in the $l$-th channel. $p$ is the three-momentum of the meson in the center of mass frame of the meson-meson system,
\begin{eqnarray}
p = \frac{\lambda^{1/2}(s,m^2_1,m^2_2)}{2 \sqrt{s}},
\label{Eq:ThreeMomentum}
\end{eqnarray}
with the K$\ddot{\text{a}}$llen function $\lambda(x,y,z) = x^2 + y^2 + z^2 - 2xy - 2yz - 2zx$.

With the isospin doublets ($D^+$, $-D^0$), ($\bar{D}^0$, $D^-$), we have,
\begin{eqnarray}
\left| D^+D^- \right\rangle &=& \frac{1}{\sqrt{2}}\left| D\bar{D}, I=0, I_3=0\right\rangle \nonumber \\
&& + \frac{1}{\sqrt{2}}\left| D\bar{D}, I=1, I_3=0\right\rangle, \\
\left| D^0\bar{D}^0 \right\rangle &=& \frac{1}{\sqrt{2}}\left| D\bar{D}, I=0, I_3=0\right\rangle  \nonumber \\
&& - \frac{1}{\sqrt{2}}\left| D\bar{D}, I=1, I_3=0\right\rangle.
\end{eqnarray}

Taking the averaged mass of $D$ meson in  Eqs.(\ref{Eq:tD0}) and (\ref{Eq:tDp}), it is easy to find that only the isospin $I=0$ component of the $D\bar{D}$ has the contribution to the $\Lambda_b\to \Lambda D\bar{D}$ process,
\begin{eqnarray}
&& G_{D^{+} D^{-}} t_{D^{+} D^{-} \rightarrow D^0 \bar{D}^0} + G_{D^0 \bar{D}^0} t_{D^0 \bar{D}^0 \rightarrow D^0 \bar{D}^0} \nonumber \\
&&= G_{D\bar{D}}  t^{I=0}_{D\bar{D}\to D\bar{D}}, \\
&& G_{D^{+} D^{-}} t_{D^{+} D^{-} \rightarrow D^{+} D^{-}}  + G_{D^0 \bar{D}^0} t_{D^0 \bar{D}^0 \rightarrow D^{+} D^{-}}  \nonumber \\
&&= G_{D\bar{D}}  t^{I=0}_{D\bar{D}\to D\bar{D}} .
\end{eqnarray}

The scattering matrices $t_{i \rightarrow j}$ in Eqs.~(\ref{Eq:tD0}) and (\ref{Eq:tDp}) are obtained by solving the Bethe-Salpeter equation in coupled channels,
\begin{eqnarray}
t=[1-VG]^{-1}V,
\label{Eq:BSequation}
\end{eqnarray}
where the elements of the diagonal matrix $G$ is the loop function of Eq.~(\ref{Eq:LoopFuntion}), and the matrix element $V_{i,j}$ are the transition potential of the $i$-th channel to the $j$-channel.
The transition potentials $V_{i,j}$($i,j = D^0 \bar{D}^0, D^{+} D^{-}, D_s^{+} D_s^{-}$) are tabulated in the Appendix A of Ref.~\cite{Gamermann:2006nm}. We introduce the potentials of $\eta \eta \rightarrow D^0 \bar{D}^0$ and $\eta \eta \rightarrow D^+ D^-$ with a dimensionless strength $a=50$ to give the width of the $D \bar{D}$ bound state, and the transition potentials of $\eta \eta \rightarrow \eta \eta$ and $\eta \eta \rightarrow D_s^{+} D_s^{-}$ are not relevant and are taken as zero~\cite{Wang:2019evy, Xiao:2012iq, Dai:2015bcc, Wang:2020elp}. Both the $G_l$ and $t_{i\to j}$ in Eqs.~(\ref{Eq:tD0}) and (\ref{Eq:tDp})  are the functions of the $D\bar{D}$ invariant mass $M_{D\bar{D}}$.

\begin{figure}[tbhp]
\begin{center}
\includegraphics[scale=0.7]{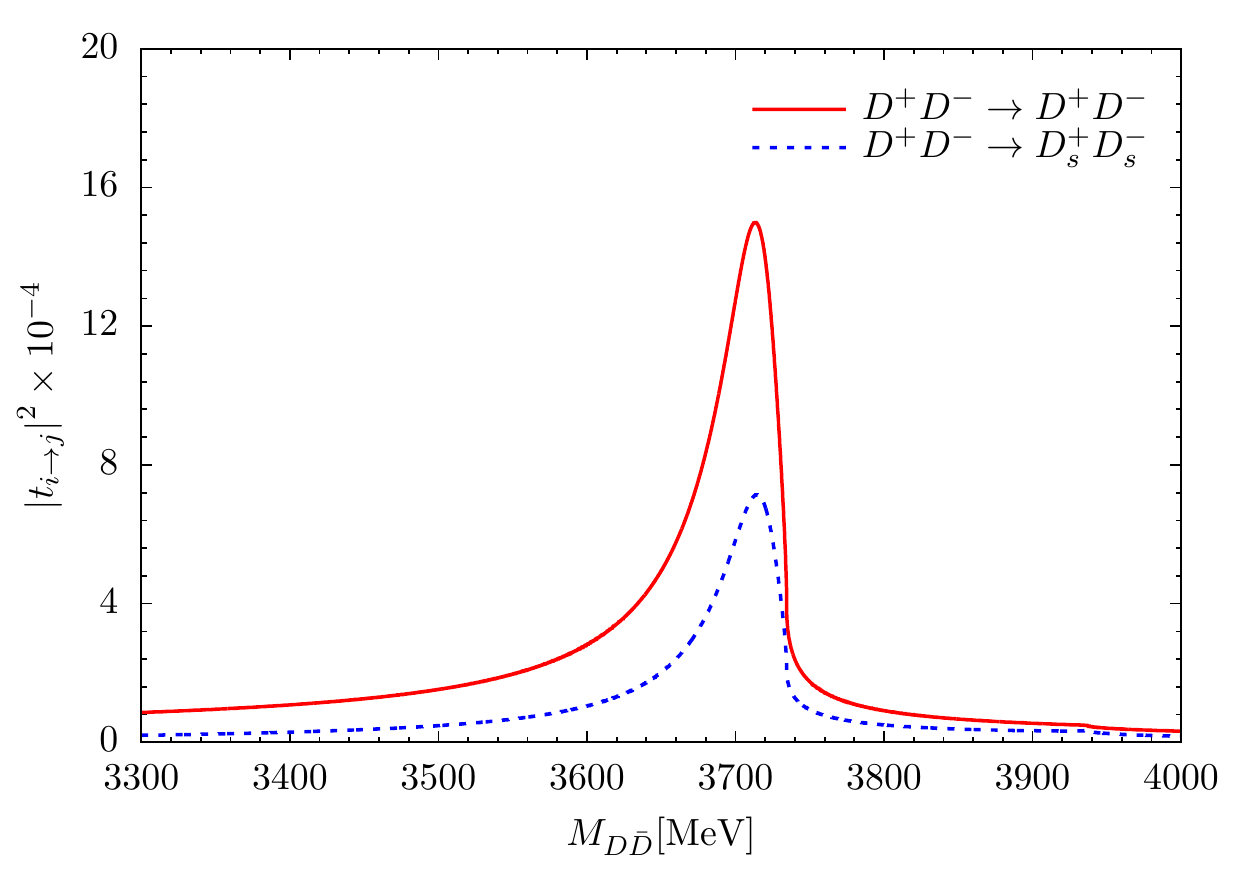}
\end{center}
%\vspace{-0.7cm}
\caption{The modulus squared of the transition amplitudes $|t_{D^+D^-\to D^+D^-}|^2$ and  $|t_{D^+D^-\to D^+_s D^-_s}|^2$ calculated with Eq.~(\ref{Eq:BSequation}).}
\label{Fig:amp}
\end{figure}

The obtained modulus squared of the transition amplitude $|t_{D^+D^-\to D^+D^-}|^2$ and $|t_{D^+D^-\to D^+_s D^-_s}|^2$ are shown in Fig.~\ref{Fig:amp}, and one can find a peak around 3720~MeV, which could be associated to the $D\bar{D}$ bound state. On the other hand, from Fig.~\ref{Fig:amp}, the $|t_{D^+ D^-\to D^+ D^-}|^2$ is two times larger than  $|t_{D^+D^-\to D^+_s D^-_s}|^2$, which indicates that the $X(3700)$ state coups mostly to $D\bar{D}$ channel.

\begin{figure}[tbhp]
\begin{center}
\includegraphics[scale = 0.9]{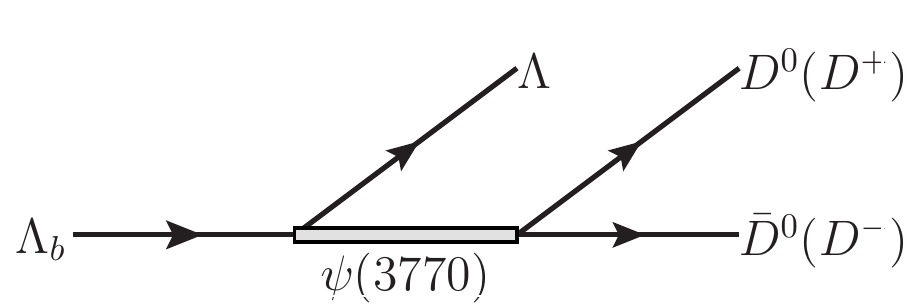}
\end{center}
%\vspace{-0.7cm}
\caption{The microscopic diagram for the decays $\Lambda_b \rightarrow \Lambda D^0 \bar{D}^0$ and $\Lambda_b \rightarrow \Lambda D^{+} D^{-}$.}
\label{Fig:3770}
\end{figure}

In addition, we also take into account  the decays $\Lambda_b \rightarrow \Lambda D^0 \bar{D}^0$ and $\Lambda_b \rightarrow \Lambda D^{+} D^{-}$ via the intermediate resonance $\psi(3770)$, which is depicted in Fig.~\ref{Fig:3770}.
The amplitude can be written as,
\begin{eqnarray}
t^{p \text{-wave}} = \frac{\beta V_p\times M_{\psi(3770)} \tilde{p}_D}{M_{D \bar{D}}^2 - M_{\psi(3770)}^2 +i M_{\psi(3770)} \tilde{\Gamma}_{\psi(3770)}},
\label{Eq:tPwave}
\end{eqnarray}
where the normalization factor $V_p$ is the same as the one in Eqs.~(\ref{Eq:tD0}) and (\ref{Eq:tDp}), and we introduce the parameter $\beta$ to account for the relative weight of the $\psi(3770)$ strength with respect to the $s$-wave contribution of Eqs.~(\ref{Eq:tD0}) and (\ref{Eq:tDp}). $\tilde{p}_D$ is the momentum of the $D^0$ (or $D^+$) in the rest frame of the $D^0 \bar{D}^0$ (or $D^{+} D^{-}$) system,
\begin{eqnarray}
\tilde{p}_D &=& \frac{\lambda^{1/2} \left( {M^2_{D \bar{D}}}, {M^2_{D}}, {M^2_{\bar{D}}} \right)}{2M_{D \bar{D}}}.
\label{Eq:DMomentum}
\end{eqnarray}

We take the width for $\psi(3770)$ energy dependent, which is given by,
\begin{eqnarray}
\tilde{\Gamma}_{\psi(3770)} = \Gamma_{\psi(3770)} \times \frac{\sqrt{M_{D \bar{D}}^2 - 4M_D^2}}{\sqrt{M_{\psi(3770)}^2 - 4M_D^2}},
\label{Eq:GammaTilde}
\end{eqnarray}
with $M_{\psi(3770)}=3773.7$~MeV, $\Gamma_{\psi(3770)}=27.2$~MeV, and $M_D=(M_{D^+}+M_{D^0})/2=1867.24$~MeV~\cite{Zyla:2020zbs}.

With the amplitudes of Eqs.~(\ref{Eq:tD0}), (\ref{Eq:tDp}) and (\ref{Eq:tPwave}), we can write the differential decay width for the decays $\Lambda_b \rightarrow \Lambda D^0 \bar{D}^0$ and $\Lambda_b \rightarrow \Lambda D^{+} D^{-}$,

\begin{eqnarray}
\frac{\mathrm{d} \Gamma }{\mathrm{d} {M_{D \bar{D}}}} \! = \! \frac{\tilde{p}_D p_{\Lambda} M_{\Lambda} M_{\Lambda_b}}{(2\pi)^3 M_{\Lambda_b}^2} \left[ {\left| t^{s \text{-wave}} \right|}^2 + {\left| t^{p \text{-wave}} \right|}^2 \right] ,
\label{Eq:DifferentialWidth}
\end{eqnarray}
with
\begin{eqnarray}
p_{\Lambda} &=& \frac{\lambda^{1/2} \left( {M_{\Lambda_b}}^2, {M_{\Lambda}}^2, {M_{D \bar{D}}}^2 \right)}{2M_{\Lambda_b}}.
\label{Eq:LambdaMomentum}
\end{eqnarray}

%%%%%%%%%%%%%%%%%%%%%%
\section{Numerical results and discussion}   \label{sec:Results}

In our model, we have three free parameters, the global normalization $V_p$,  the color factor $C$, and $\beta$. $V_p$ is a global factor and its value does not affect the shapes of the $D^0 \bar{D}^0$ and $D^+ D^-$ invariant mass distributions. $\beta$ represents the relative weight of the $\psi(3770)$ strength with respect to the one of $s$-wave, and we take its value $\beta=0.15$ to give the contributions from the $s$-wave $D\bar{D}$ interaction and the $\psi(3770)$ with the same order of magnitude.  Next, we first show the results with the color factor $C=3$ and $V_p=1$, and will present the results for different values of $C$ and $\beta$.

\begin{figure}[tbhp]
\begin{center}
\includegraphics[scale=0.6]{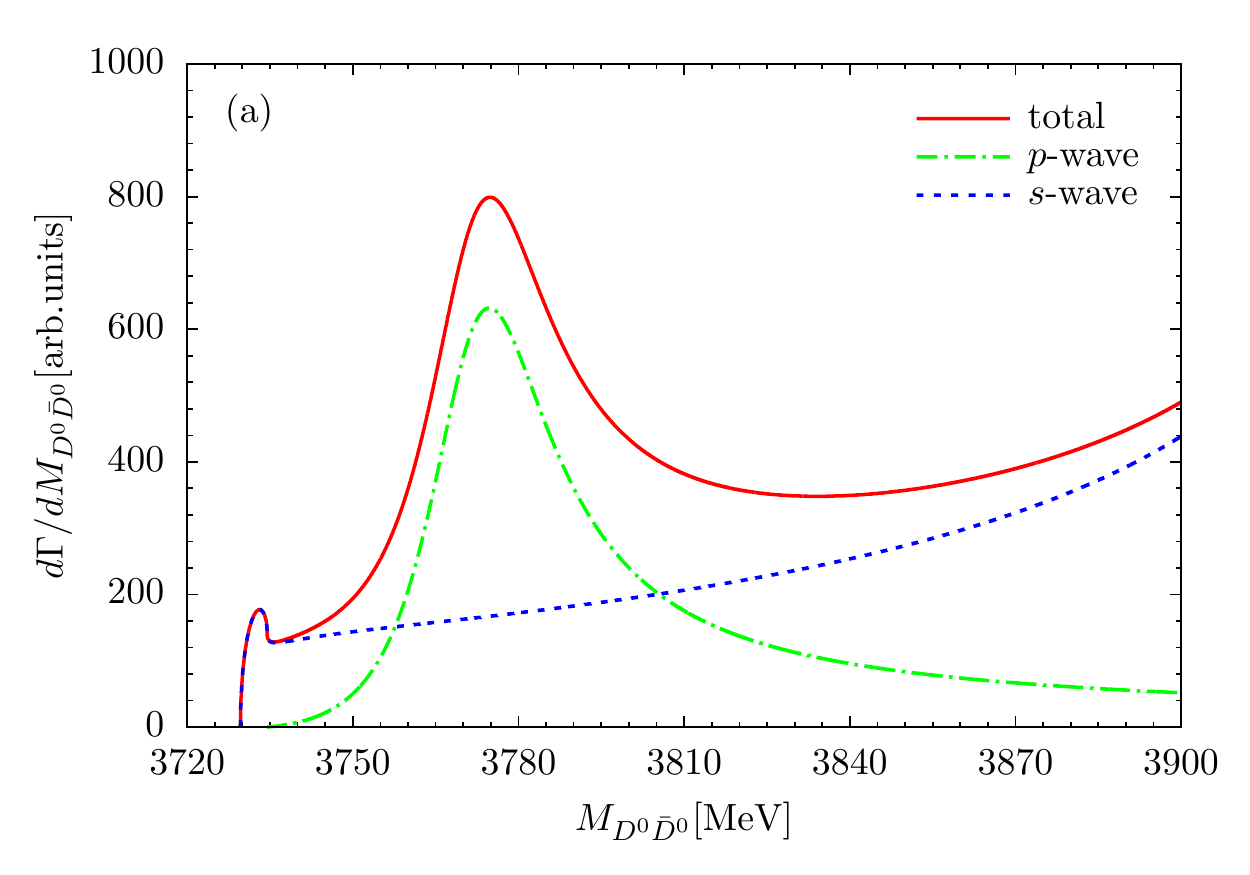}
\includegraphics[scale=0.6]{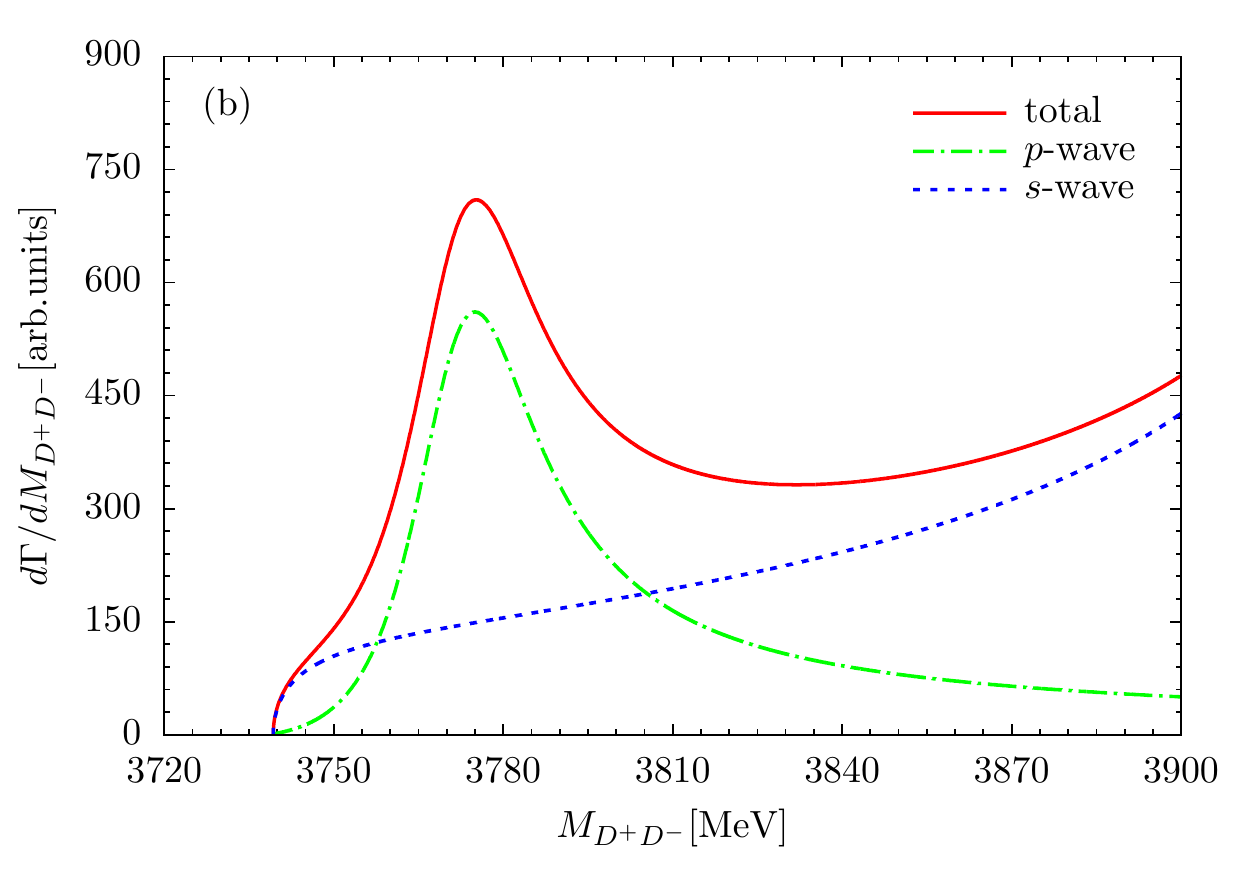}
\end{center}
%\vspace{-0.7cm}
\caption{The ${D^0 \bar{D}^0}$ (a) and  ${D^+ D^-}$ (b) invariant mass distributions of the processes   $\Lambda_b \rightarrow \Lambda D^0 \bar{D}^0$ and $\Lambda_b \rightarrow \Lambda D^+ {D}^-$. The blue dashed curve shows the contribution from the meson-meson interaction in $s$-wave, the green dash-dotted curve corresponds to the results for the intermediate meson $\psi(3770)$, and the red solid curve shows the total contributions.}
\label{Fig:ds_DD}
\end{figure}

We show the $D^0 \bar{D}^0$ and $D^+ D^-$ invariant mass distributions in Fig.~\ref{Fig:ds_DD}. One can find a clear enhancement near the $D^0 \bar{D}^0$ threshold in the $D^0 \bar{D}^0$ invariant mass distribution of the $\Lambda_b \to \Lambda D^0\bar{D}^0$,  due to the presence of the $X(3700)$ resonance below the $D\bar{D}$ threshold. The enhancement structure near the threshold is a little weaker for the $D^+{D}^-$ invariant mass distribution of the  $\Lambda_b \to \Lambda D^+D^-$, because the $D^{+} D^{-}$ threshold is higher than the $D^0 \bar{D}^0$ one and farther away from the peak of $X(3700)$.

\begin{figure}[tbhp]
\begin{center}
\includegraphics[scale=0.6]{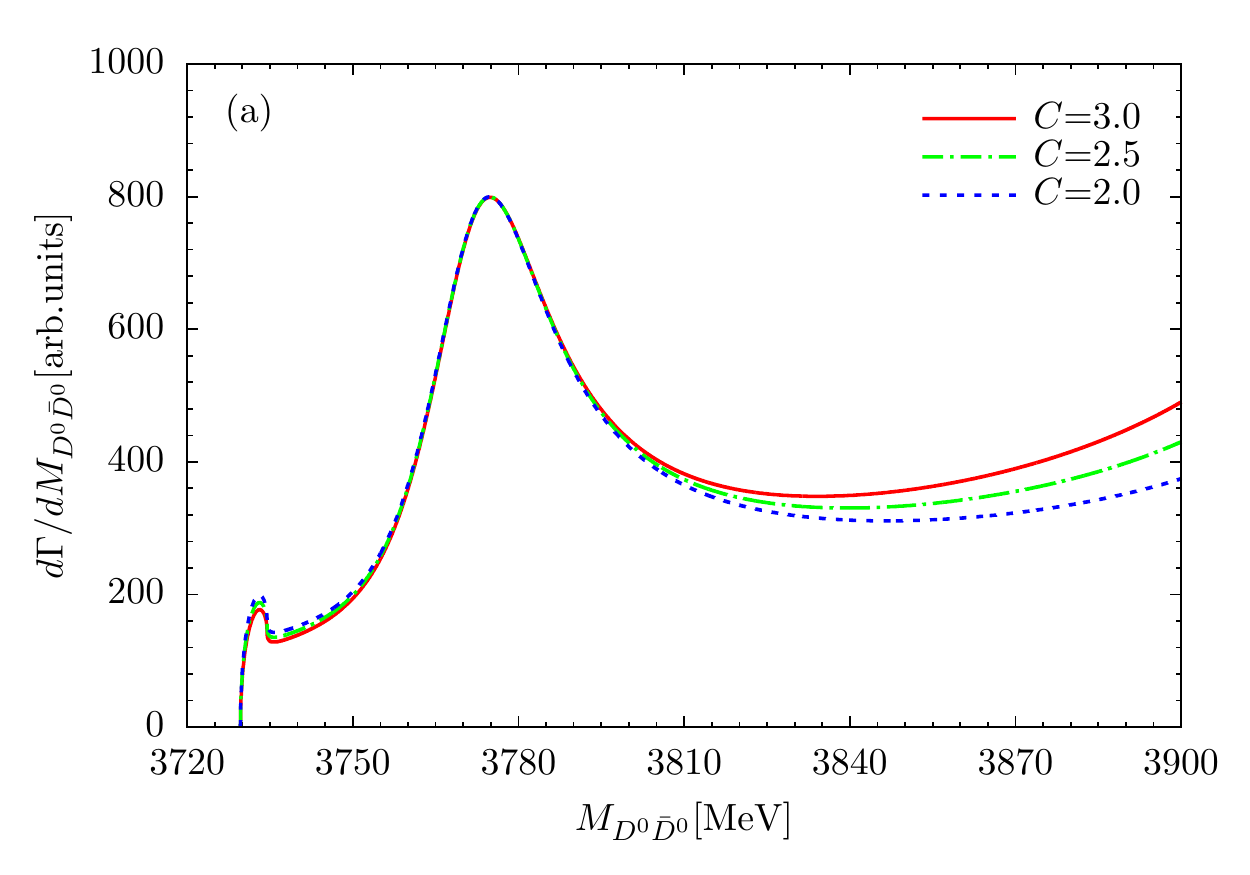}
\includegraphics[scale=0.6]{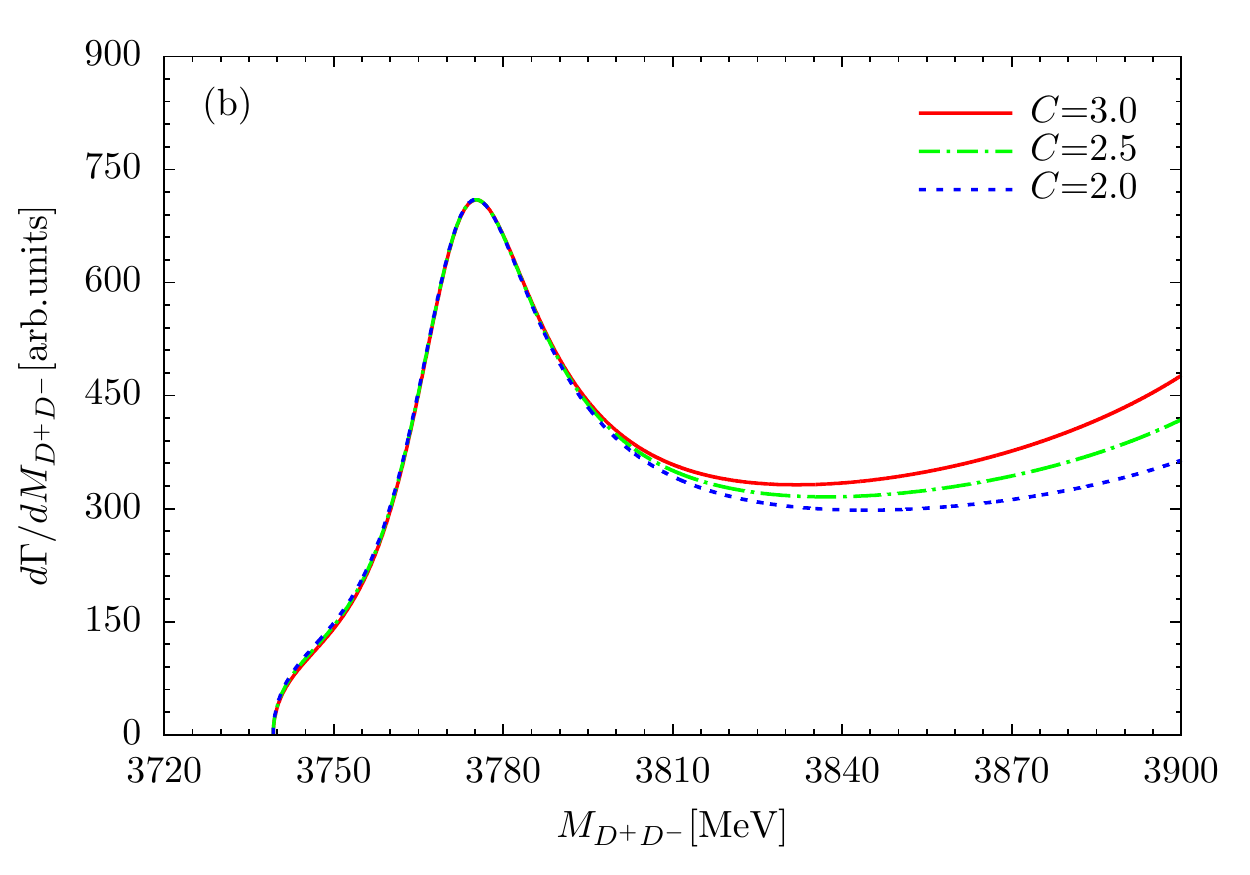}
\end{center}
%\vspace{-0.7cm}
\caption{The ${D^0 \bar{D}^0}$ (a) and  ${D^+ D^-}$ (b) invariant mass distributions of the processes   $\Lambda_b \rightarrow \Lambda D^0 \bar{D}^0$ and $\Lambda_b \rightarrow \Lambda D^+ {D}^-$ with different values of $C$ = 3.0, 2.5, 2.0.}
\label{Fig:ds_C}
\end{figure}

In Fig.~\ref{Fig:ds_C}, we show the $D^0\bar{D}^0$ and  $D^+D^-$  invariant mass distributions with the different values of color factor $C=3.0, 2.5,2.0$. One can find that both mass distributions near the threshold do not change too much, since the value of color factor $C$ only affects the contribution from the $D_s^+D^-_s$ loop of Fig.~\ref{Fig:Feynman}(b), which is smaller than the contributions from the $D^+D^-$ and $D^0\bar{D}^0$.

\begin{figure}[tbhp]
\begin{center}
\includegraphics[scale=0.6]{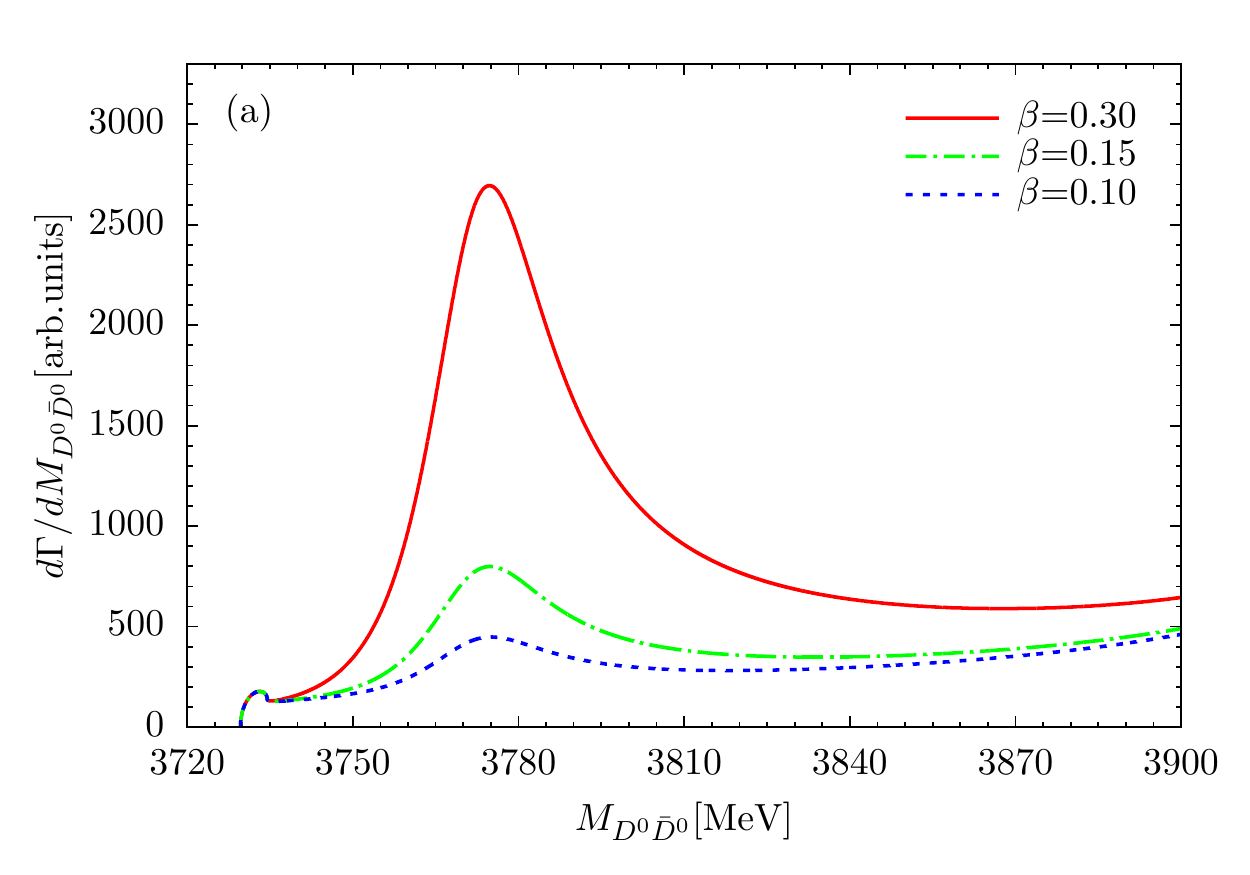}
\includegraphics[scale=0.6]{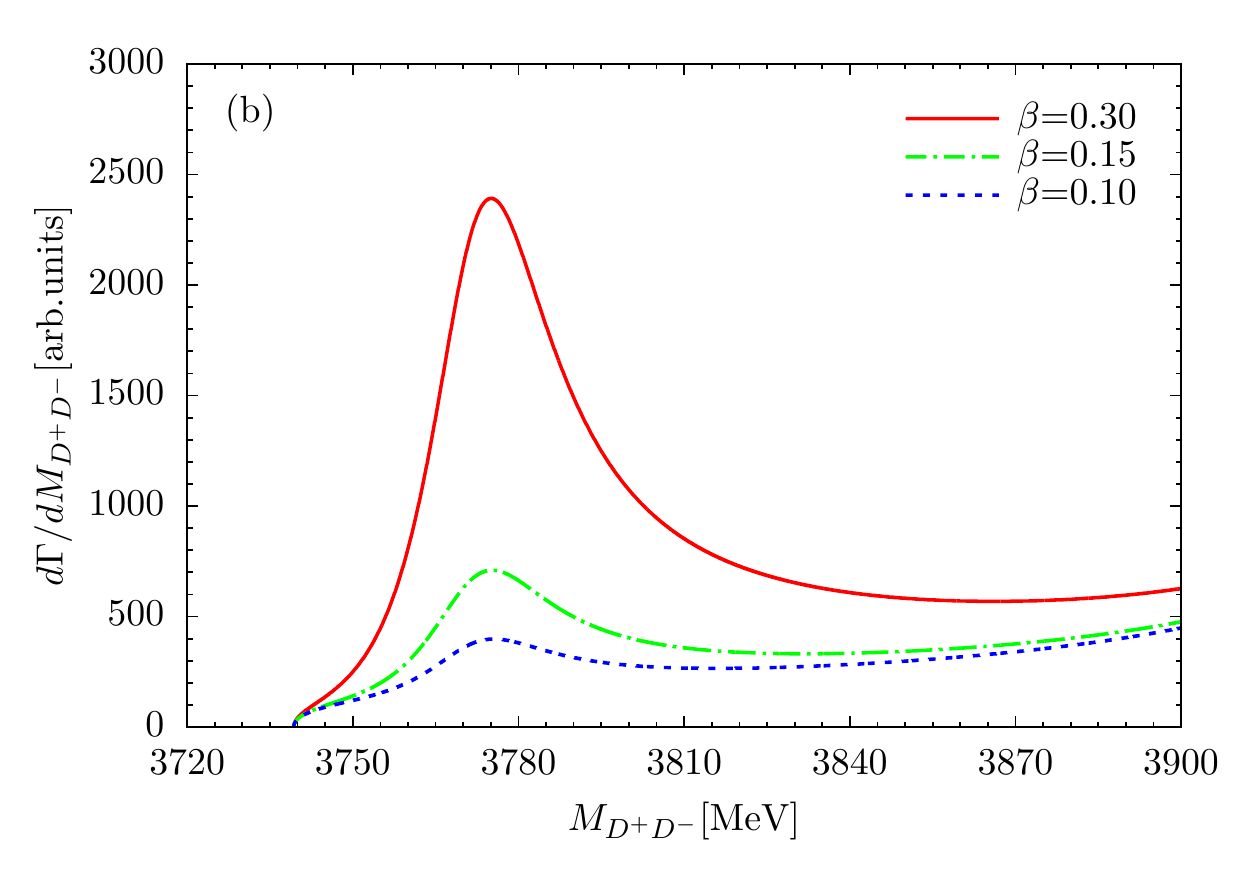}
\end{center}
%\vspace{-0.7cm}
\caption{The ${D^0 \bar{D}^0}$ (a) and  ${D^+ D^-}$ (b) invariant mass distributions of the processes   $\Lambda_b \rightarrow \Lambda D^0 \bar{D}^0$ and $\Lambda_b \rightarrow \Lambda D^+ {D}^-$ with different values of $\beta$ =0.30, 0.15, 0.10.}
\label{Fig:ds_beta}
\end{figure}

We also present our results for the different values of $\beta=0.30, 0.15, 0.10$ in Fig.~\ref{Fig:ds_beta}. One can see that the enhancement near the threshold will be identified difficultly for the larger value of $\beta$. Indeed, the
$\psi(3770)$ would provide the dominant contribution for the $\Lambda_b\to \Lambda D\bar{D}$ process, however, it is still
expected to find an enhancement near the $D\bar{D}$ threshold, especially the $D^0\bar{D}^0$ one, if the $D\bar{D}$ bound state do exist, as predicted in Refs.~\cite{Gamermann:2006nm,Prelovsek:2020eiw}.
Furthermore,  since the $\psi(3770)$ state couples to $D\bar{D}$ in $p$-wave, the partial wave analysis of this reaction would be helpful to test the existence of the $D\bar{D}$ bound state.

At present, the  LHCb Collaboration has accumulated a large number of $\Lambda_b$ events, thus, we would like to call the attention of the experimentalists to measure the $\Lambda_b \rightarrow \Lambda D \bar{D}$ decay, which should be useful to confirm the existence of $X(3700)$ and to understand its nature.

%%%%%%%%%%%%%%%%%%%%%%
\section{Conclusions} \label{sec:Conclusions}

The study of the charmonium-like states is crucial to understand the heavy-hadron heavy-hadron interactions, and also the internal structures of the hidden-charm states. One $D\bar{D}$ bound state around 3700~MeV was predicted within the coupled channel unitary approach~\cite{Gamermann:2006nm}, and also the lattice investigation of the $D\bar{D}$ and $D_s\bar{D}_s$ scattering~\cite{Wang:2020elp}. Although our previous studies on the $e^+e^-\to J/\psi D\bar{D}$ and $\gamma\gamma \to D\bar{D}$ data support the existence of the $D\bar{D}$ bound state, the other possibilities cannot be discarded due to the present quality of the experimental data~\cite{Wang:2019evy,Wang:2020elp}. Investigating the processes involving the $s$-wave $D\bar{D}$ system could provide the information about the existence of the $D\bar{D}$ bound state.

In this paper, we have investigated the processes  $\Lambda_b \rightarrow \Lambda D^0 \bar{D}^0$ and $\Lambda_b \rightarrow \Lambda D^+ D^-$  within the coupled channel unitary approach, by taking into account the $s$-wave meson-meson interactions and the contribution from the intermediate resonance $\psi(3770)$. The $D^0 \bar{D}^0$ and $D^+ D^-$ invariant mass distributions in the $\Lambda_b \to \Lambda D \bar{D}$ reaction are investigated, and our results show an enhancement structure near the $D\bar{D}$ threshold, which should be the reflection of the $D\bar{D}$ bound state. Therefore, we strongly encourage our experimental colleagues to measure the $\Lambda_b \rightarrow \Lambda D \bar{D}$ process, which would be crucial to confirm the existence the $X(3700)$ resonance, and to  understand the heavy-hadron heavy-hadron interactions.

%%%%%%%%%%%%%%%%%%%%%%
\begin{acknowledgments}

This work is partly supported by the National Natural Science Foundation of China under Grants Nos. 11947089, 12075288, 11735003, and 11961141012.  It is also supported by the Key Research Projects of Henan Higher Education Institutions under No. 20A140027, Training Plan for Young Key Teachers in Higher Schools in Henan Province (2020GGJS017), the Academic Improvement Project of Zhengzhou University, the Fundamental Research Cultivation Fund for Young Teachers of Zhengzhou University (JC202041042), and the Youth Innovation Promotion Association CAS (2016367).

\end{acknowledgments}

%%%%%%%%%%%%%%%%%%%%%%
%\bibliographystyle{unsrt}
%\bibliography{spectral}

%\bibliographystyle{unsrt}
%\bibliography{ref}

\begin{thebibliography}{99}

%\cite{GellMann:1964nj}
\bibitem{GellMann:1964nj}
M.~Gell-Mann,
A Schematic Model of Baryons and Mesons,
Phys. Lett. \textbf{8} (1964), 214-215.
%doi:10.1016/S0031-9163(64)92001-3
%3359 citations counted in INSPIRE as of 10 Jan 2021

%\cite{Zweig:1964jf}
\bibitem{Zweig:1964jf}
G.~Zweig,
An $SU(3)$ model for strong interaction symmetry and its breaking. Version 2,
CERN-TH-412.
Developments in the Quark Theory of Hadrons, Volume 1. Edited by D. Lichtenberg and S. Rosen. pp. 22-101
%556 citations counted in INSPIRE as of 21 Jan 2021

\bibitem{Choi:2003ue}
S.~K.~Choi \textit{et al.} [Belle],
Observation of a narrow charmonium - like state in exclusive $B^{\pm}\to K^{\pm} \pi^+ \pi^- J/\psi$ decays,
Phys. Rev. Lett. \textbf{91} (2003), 262001.
%doi:10.1103/PhysRevLett.91.262001
%[arXiv:hep-ex/0309032 [hep-ex]].
%1884 citations counted in INSPIRE as of 10 Jan 2021


%\cite{Zyla:2020zbs}
\bibitem{Zyla:2020zbs}
P.~A.~Zyla \textit{et al.} [Particle Data Group],
Review of Particle Physics,
PTEP \textbf{2020} (2020) no.8, 083C01.
%doi:10.1093/ptep/ptaa104
%643 citations counted in INSPIRE as of 21 Jan 2021


%\cite{Brambilla:2019esw}
\bibitem{Brambilla:2019esw}
N.~Brambilla, S.~Eidelman, C.~Hanhart, A.~Nefediev, C.~P.~Shen, C.~E.~Thomas, A.~Vairo and C.~Z.~Yuan,
The $XYZ$ states: experimental and theoretical status and perspectives,
Phys. Rept. \textbf{873} (2020), 1-154.
%doi:10.1016/j.physrep.2020.05.001
%[arXiv:1907.07583 [hep-ex]].
%159 citations counted in INSPIRE as of 29 Jan 2021

%\cite{Olsen:2017bmm}
\bibitem{Olsen:2017bmm}
S.~L.~Olsen, T.~Skwarnicki and D.~Zieminska,
Nonstandard heavy mesons and baryons: Experimental evidence,
Rev. Mod. Phys. \textbf{90} (2018) no.1, 015003.
%doi:10.1103/RevModPhys.90.015003
%[arXiv:1708.04012 [hep-ph]].
%325 citations counted in INSPIRE as of 29 Jan 2021



%\cite{Chen:2016qju}
\bibitem{Chen:2016qju}
H.~X.~Chen, W.~Chen, X.~Liu and S.~L.~Zhu,
The hidden-charm pentaquark and tetraquark states,
Phys. Rept. \textbf{639} (2016), 1-121.
%doi:10.1016/j.physrep.2016.05.004
%[arXiv:1601.02092 [hep-ph]].
%622 citations counted in INSPIRE as of 21 Jan 2021

%\cite{Liu:2019zoy}
\bibitem{Liu:2019zoy}
Y.~R.~Liu, H.~X.~Chen, W.~Chen, X.~Liu and S.~L.~Zhu,
Pentaquark and Tetraquark states,
Prog. Part. Nucl. Phys. \textbf{107} (2019), 237-320.
%doi:10.1016/j.ppnp.2019.04.003
%[arXiv:1903.11976 [hep-ph]].
%192 citations counted in INSPIRE as of 21 Jan 2021

%\cite{Hosaka:2016pey}
\bibitem{Hosaka:2016pey}
A.~Hosaka, T.~Iijima, K.~Miyabayashi, Y.~Sakai and S.~Yasui,
Exotic hadrons with heavy flavors: $X$, $Y$, $Z$, and related states,
PTEP \textbf{2016} (2016) no.6, 062C01.
%doi:10.1093/ptep/ptw045
%[arXiv:1603.09229 [hep-ph]].
%129 citations counted in INSPIRE as of 21 Jan 2021

%\cite{Guo:2017jvc}
\bibitem{Guo:2017jvc}
F.~K.~Guo, C.~Hanhart, U.~G.~Mei\ss{}ner, Q.~Wang, Q.~Zhao and B.~S.~Zou,
Hadronic molecules,
Rev. Mod. Phys. \textbf{90} (2018) no.1, 015004.
%doi:10.1103/RevModPhys.90.015004
%[arXiv:1705.00141 [hep-ph]].
%496 citations counted in INSPIRE as of 21 Jan 2021

%\cite{Hao:2019fjg}
\bibitem{Hao:2019fjg}
W.~Hao, G.~Y.~Wang, E.~Wang, G.~N.~Li and D.~M.~Li,
Canonical interpretation of the $X(4140)$ state within the $^3P_0$ model,
Eur. Phys. J. C \textbf{80} (2020) no.7, 626.
%doi:10.1140/epjc/s10052-020-8187-0
%[arXiv:1909.13099 [hep-ph]].
%3 citations counted in INSPIRE as of 07 Feb 2021


%\cite{Dong:2020hxe}
\bibitem{Dong:2020hxe}
X.~K.~Dong, F.~K.~Guo and B.~S.~Zou,
Why there are many threshold structures in hadron spectrum with heavy quarks,
[arXiv:2011.14517 [hep-ph]].
%3 citations counted in INSPIRE as of 10 Jan 2021


%\cite{Wang:2018djr}
\bibitem{Wang:2018djr}
E.~Wang, J.~J.~Xie, L.~S.~Geng and E.~Oset,
The $X(4140)$ and $X(4160)$ resonances in the $e^+e^-\to \gamma J/\psi \phi $ reaction,
Chin. Phys. C \textbf{43} (2019) no.11, 113101.
%doi:10.1088/1674-1137/43/11/113101
%[arXiv:1806.05113 [hep-ph]].
%8 citations counted in INSPIRE as of 07 Feb 2021

%\cite{Wang:2017mrt}
\bibitem{Wang:2017mrt}
E.~Wang, J.~J.~Xie, L.~S.~Geng and E.~Oset,
Analysis of the $B^+\to J/\psi \phi K^+$ data at low $J/\psi \phi$ invariant masses and the $X(4140)$ and $X(4160)$ resonances,
Phys. Rev. D \textbf{97} (2018) no.1, 014017.
%doi:10.1103/PhysRevD.97.014017
%[arXiv:1710.02061 [hep-ph]].
%21 citations counted in INSPIRE as of 07 Feb 2021



%\cite{Gamermann:2006nm}
\bibitem{Gamermann:2006nm}
D.~Gamermann, E.~Oset, D.~Strottman and M.~J.~Vicente Vacas,
Dynamically generated open and hidden charm meson systems,
Phys. Rev. D \textbf{76} (2007), 074016.
%doi:10.1103/PhysRevD.76.074016
%[arXiv:hep-ph/0612179 [hep-ph]].
%250 citations counted in INSPIRE as of 30 Dec 2020



%\cite{Dai:2015bcc}
\bibitem{Dai:2015bcc}
L.~R.~Dai, J.~J.~Xie and E.~Oset,
$B^0 \rightarrow D^0 \bar{D}^0 K^0$ , $B^+ \rightarrow D^0 \bar{D}^0 K^+$ , and the scalar $D \bar{D}$ bound state,
Eur. Phys. J. C \textbf{76} (2016) no.3, 121.
%doi:10.1140/epjc/s10052-016-3946-7
%[arXiv:1512.04048 [hep-ph]].
%10 citations counted in INSPIRE as of 03 Jan 2021


%\cite{Xiao:2012iq}
\bibitem{Xiao:2012iq}
C.~W.~Xiao and E.~Oset,
Three methods to detect the predicted $D \bar{D}$ scalar meson $X(3700)$,
Eur. Phys. J. A \textbf{49} (2013), 52.
%doi:10.1140/epja/i2013-13052-5
%[arXiv:1211.1862 [hep-ph]].

%\cite{Wang:2019evy}
\bibitem{Wang:2019evy}
E.~Wang, W.~H.~Liang and E.~Oset,
Analysis of the ${e^+ e^- \to J/\psi D\bar D}$ reaction close to the threshold concerning claims of a $\chi_{c0}(2P)$ state,
Eur. Phys. J. A \textbf{57} (2021), 38.
%[arXiv:1902.06461 [hep-ph]].
%7 citations counted in INSPIRE as of 03 Jan 2021

 %\cite{Gamermann:2007mu}
\bibitem{Gamermann:2007mu}
D.~Gamermann and E.~Oset,
Hidden charm dynamically generated resonances and the $e^+ e^-\to J / \psi D \bar{D}, J / \psi D \bar{D}^*$ reactions,
Eur. Phys. J. A \textbf{36} (2008), 189-194.
%doi:10.1140/epja/i2007-10580-5
%[arXiv:0712.1758 [hep-ph]].
%60 citations counted in INSPIRE as of 13 Jan 2021



%\cite{Abe:2007sya}
\bibitem{Abe:2007sya}
P.~Pakhlov \textit{et al.} [Belle],
Production of New Charmoniumlike States in $e^+ e^- \to J/\psi D^{(*)} \bar{D}^{(*)}$ at $s^{1/2}\sim 10$ GeV,
Phys. Rev. Lett. \textbf{100} (2008), 202001.
%doi:10.1103/PhysRevLett.100.202001
%[arXiv:0708.3812 [hep-ex]].
%237 citations counted in INSPIRE as of 13 Jan 2021

%\cite{Chilikin:2017evr}
\bibitem{Chilikin:2017evr}
K.~Chilikin \textit{et al.} [Belle],
Observation of an alternative $\chi_{c0}(2P)$ candidate in $e^+ e^- \rightarrow J/\psi D \bar{D}$,
Phys. Rev. D \textbf{95} (2017), 112003.
%doi:10.1103/PhysRevD.95.112003
%[arXiv:1704.01872 [hep-ex]].
%62 citations counted in INSPIRE as of 13 Jan 2021


%\cite{Uehara:2005qd}
\bibitem{Uehara:2005qd}
  S.~Uehara {\it et al.} [Belle Collaboration],
  Observation of a $\chi^\prime_{c2}$ candidate in $\gamma \gamma \to D \bar D$ production at BELLE,
  Phys.\ Rev.\ Lett.\  {\bf 96}, 082003 (2006).
%  doi:10.1103/PhysRevLett.96.082003
%  [hep-ex/0512035].
  %%CITATION = doi:10.1103/PhysRevLett.96.082003;%%
  %295 citations counted in INSPIRE as of 15 Jan 2020

%\cite{Aubert:2010ab}
\bibitem{Aubert:2010ab}
  B.~Aubert {\it et al.} [BaBar Collaboration],
  Observation of the $\chi_{c2}(2P)$ Meson in the Reaction $\gamma \gamma \to D \bar{D}$ at {BaBar},
  Phys.\ Rev.\ D {\bf 81}, 092003 (2010).
%  doi:10.1103/PhysRevD.81.092003
%  [arXiv:1002.0281 [hep-ex]].
  %%CITATION = doi:10.1103/PhysRevD.81.092003;%%
  %93 citations counted in INSPIRE as of 15 Jan 2020

%\cite{Wang:2020elp}
\bibitem{Wang:2020elp}
E.~Wang, H.~S.~Li, W.~H.~Liang and E.~Oset,
Analysis of the ${\gamma\gamma \to D\bar D}$ reaction and the $D\bar{D}$ bound state,
[arXiv:2010.15431 [hep-ph]]. Accepted by PRD.
%3 citations counted in INSPIRE as of 03 Jan 2021

%\cite{Prelovsek:2020eiw}
\bibitem{Prelovsek:2020eiw}
S.~Prelovsek, S.~Collins, D.~Mohler, M.~Padmanath and S.~Piemonte,
Charmonium-like resonances with $J^{PC}=0^{++},2^{++}$ in coupled $D\bar D$, $D_s\bar D_s$ scattering on the lattice,
[arXiv:2011.02542 [hep-lat]].
%1 citations counted in INSPIRE as of 04 Dec 2020

%\cite{Oset:2016lyh}
\bibitem{Oset:2016lyh}
E.~Oset, W.~H.~Liang, M.~Bayar, J.~J.~Xie, L.~R.~Dai, M.~Albaladejo, M.~Nielsen, T.~Sekihara, F.~Navarra and L.~Roca, \textit{et al.}
Weak decays of heavy hadrons into dynamically generated resonances,
Int. J. Mod. Phys. E \textbf{25} (2016), 1630001.
%doi:10.1142/S0218301316300010
%[arXiv:1601.03972 [hep-ph]].
%73 citations counted in INSPIRE as of 29 Jan 2021


%\cite{Aaij:2019dvk}
\bibitem{Aaij:2019dvk}
R.~Aaij \textit{et al.} [LHCb],
Measurement of the ratio of branching fractions of the decays $\Lambda^0_b\to\psi(2S) \Lambda$ and $\Lambda^0_b\!\to J/\psi \Lambda$,
JHEP \textbf{03} (2019), 126.
%doi:10.1007/JHEP03(2019)126
%[arXiv:1902.02092 [hep-ex]].
%6 citations counted in INSPIRE as of 29 Jan 2021

%\cite{Abazov:2011wt}
\bibitem{Abazov:2011wt}
V.~M.~Abazov \textit{et al.} [D0],
Measurement of the production fraction times branching fraction $f(b\to\Lambda_{b})\cdot \mathcal{B}(\Lambda_{b}\to J/\psi \Lambda)$,
Phys. Rev. D \textbf{84} (2011), 031102.
%doi:10.1103/PhysRevD.84.031102
%[arXiv:1105.0690 [hep-ex]].
%29 citations counted in INSPIRE as of 29 Jan 2021

%\cite{Aad:2015msa}
\bibitem{Aad:2015msa}
G.~Aad \textit{et al.} [ATLAS],
Measurement of the branching ratio $\Gamma(\Lambda_b^0 \rightarrow \psi(2S)\Lambda^0)/\Gamma(\Lambda_b^0 \rightarrow J/\psi\Lambda^0)$ with the ATLAS detector,
Phys. Lett. B \textbf{751} (2015), 63-80.
%doi:10.1016/j.physletb.2015.10.009
%[arXiv:1507.08202 [hep-ex]].
%24 citations counted in INSPIRE as of 29 Jan 2021




%\cite{Hsiao:2015txa}
\bibitem{Hsiao:2015txa}
Y.~K.~Hsiao, P.~Y.~Lin, L.~W.~Luo and C.~Q.~Geng,
Fragmentation fractions of two-body $b$-baryon decays,
Phys. Lett. B \textbf{751} (2015), 127-130.
%doi:10.1016/j.physletb.2015.10.013
%[arXiv:1510.01808 [hep-ph]].
%16 citations counted in INSPIRE as of 10 Jan 2021

%\cite{Aaij:2016nrq}
\bibitem{Aaij:2016nrq}
R.~Aaij \textit{et al.} [LHCb],
Observations of $\Lambda_b^0 \to \Lambda K^+\pi^-$ and $\Lambda_b^0 \to \Lambda K^+K^-$ decays and searches for other $\Lambda_b^0$ and $\Xi_b^0$ decays to $\Lambda h^+h^{\prime -}$ final states,
JHEP \textbf{05} (2016), 081.
%doi:10.1007/JHEP05(2016)081
%[arXiv:1603.00413 [hep-ex]].
%35 citations counted in INSPIRE as of 10 Jan 2021




%\cite{Wang:2020wap}
\bibitem{Wang:2020wap}
G.~Y.~Wang, M.~Y.~Duan, E.~Wang and D.~M.~Li,
Enhancement near the $\bar{p}\Lambda$ threshold in the $\chi_{c0}\to \bar{p}K^+\Lambda$ reaction,
Phys. Rev. D \textbf{102} (2020) no.3, 036003.
%doi:10.1103/PhysRevD.102.036003
%[arXiv:2003.03894 [hep-ph]].
%1 citations counted in INSPIRE as of 29 Jan 2021

%\cite{Ablikim:2006dw}
\bibitem{Ablikim:2006dw}
  M.~Ablikim {\it et al.} [BES Collaboration],
  Observation of a near-threshold enhancement in the $\omega \phi$ mass spectrum from the doubly OZI suppressed decay $J / \psi \to \gamma \omega \phi$,
  Phys.\ Rev.\ Lett.\  {\bf 96}, 162002 (2006).
%  doi:10.1103/PhysRevLett.96.162002
%  [hep-ex/0602031].
  %%CITATION = doi:10.1103/PhysRevLett.96.162002;%%
  %109 citations counted in INSPIRE as of 31 Jan 2020


%\cite{Geng:2008gx}
\bibitem{Geng:2008gx}
  L.~S.~Geng and E.~Oset,
  Vector meson-vector meson interaction in a hidden gauge unitary approach,
  Phys.\ Rev.\ D {\bf 79}, 074009 (2009).
%  doi:10.1103/PhysRevD.79.074009
%  [arXiv:0812.1199 [hep-ph]].
  %%CITATION = doi:10.1103/PhysRevD.79.074009;%%
  %145 citations counted in INSPIRE as of 31 Jan 2020

%\cite{Ablikim:2009ac}
\bibitem{Ablikim:2009ac}
  M.~Ablikim {\it et al.} [BES Collaboration],
  Study of $J/\psi$ decays into $\eta K^{*0} \bar{K}^{*0}$,
  Phys.\ Lett.\ B {\bf 685}, 27 (2010).
%  doi:10.1016/j.physletb.2010.01.063
%  [arXiv:0909.2087 [hep-ex]].
  %%CITATION = doi:10.1016/j.physletb.2010.01.063;%%
  %19 citations counted in INSPIRE as of 31 Jan 2020

%\cite{Xie:2013ula}
\bibitem{Xie:2013ula}
  J.~J.~Xie, M.~Albaladejo and E.~Oset,
  Signature of an $h_1$ state in the $J/\psi \to \eta h_1 \to \eta K^{*0}\bar{K}^{*0}$ decay,
  Phys.\ Lett.\ B {\bf 728}, 319 (2014).
%  doi:10.1016/j.physletb.2013.12.015
%  [arXiv:1306.6594 [hep-ph]].
  %%CITATION = doi:10.1016/j.physletb.2013.12.015;%%
  %18 citations counted in INSPIRE as of 31 Jan 2020


%\cite{Lu:2016roh}
\bibitem{Lu:2016roh}
J.~X.~Lu, E.~Wang, J.~J.~Xie, L.~S.~Geng and E.~Oset,
The $\Lambda_{b}\rightarrow J/\psi K^{0}\Lambda$ reaction and a hidden-charm pentaquark state with strangeness,
Phys. Rev. D \textbf{93} (2016), 094009.
%doi:10.1103/PhysRevD.93.094009
%[arXiv:1601.00075 [hep-ph]].
%39 citations counted in INSPIRE as of 29 Dec 2020


















%\cite{Wang:2015pcn}
\bibitem{Wang:2015pcn}
E.~Wang, H.~X.~Chen, L.~S.~Geng, D.~M.~Li and E.~Oset,
Hidden-charm pentaquark state in $\Lambda^0_b \to J/\psi p \pi^-$ decay,
Phys. Rev. D \textbf{93} (2016) no.9, 094001.
%doi:10.1103/PhysRevD.93.094001
%[arXiv:1512.01959 [hep-ph]].
%43 citations counted in INSPIRE as of 29 Dec 2020

%\cite{Li:2020fqp}
\bibitem{Li:2020fqp}
H.~S.~Li, L.~L.~Wei, M.~Y.~Duan, E.~Wang and D.~M.~Li,
Search for the scalar meson $a_0(980)$ in the single Cabibbo-suppressed process $\Lambda_c \to \pi^0\eta p$,
[arXiv:2009.08600 [hep-ph]].
%1 citations counted in INSPIRE as of 29 Dec 2020

%\cite{Wang:2020pem}
\bibitem{Wang:2020pem}
Z.~Wang, Y.~Y.~Wang, E.~Wang, D.~M.~Li and J.~J.~Xie,
The scalar $f_0(500)$ and $f_0(980)$ resonances and vector mesons in the single Cabibbo-suppressed decays $\Lambda_c \to p K^+K^-$ and $p\pi^+\pi^-$,
Eur. Phys. J. C \textbf{80} (2020) no.9, 842.
%doi:10.1140/epjc/s10052-020-8347-2
%[arXiv:2004.01438 [hep-ph]].
%3 citations counted in INSPIRE as of 29 Dec 2020

%\cite{Liu:2020ajv}
\bibitem{Liu:2020ajv}
W.~Y.~Liu, W.~Hao, G.~Y.~Wang, Y.~Y.~Wang, E.~Wang and D.~M.~Li,
The resonances $X(4140)$, $X(4160)$, and $P_{cs}(4459)$ in the decay of $\Lambda_b\to J/\psi\Lambda\phi$,
[arXiv:2012.01804 [hep-ph]]. Accepted by PRD.
%0 citations counted in INSPIRE as of 30 Dec 2020


%\cite{Duan:2020vye}
\bibitem{Duan:2020vye}
M.~Y.~Duan, J.~Y.~Wang, G.~Y.~Wang, E.~Wang and D.~M.~Li,
Role of scalar $a_0(980)$ in the single Cabibbo suppressed process $D^+ \rightarrow \pi ^{+} \pi ^{0} \eta $,
Eur. Phys. J. C \textbf{80} (2020) no.11, 1041.
%doi:10.1140/epjc/s10052-020-08630-3
%[arXiv:2008.10139 [hep-ph]].
%4 citations counted in INSPIRE as of 07 Feb 2021

%\cite{Zhang:2020rqr}
\bibitem{Zhang:2020rqr}
Y.~Zhang, E.~Wang, D.~M.~Li and Y.~X.~Li,
Search for the $D^*\bar{D}^*$ molecular state $Z_c(4000)$ in the reaction $B^{-} \rightarrow J/\psi \rho^0 K^{-}$,
Chin. Phys. C \textbf{44} (2020) no.9, 093107.
%doi:10.1088/1674-1137/44/9/093107
%[arXiv:2001.06624 [hep-ph]].
%3 citations counted in INSPIRE as of 07 Feb 2021

%\cite{Dai:2018nmw}
\bibitem{Dai:2018nmw}
L.~R.~Dai, G.~Y.~Wang, X.~Chen, E.~Wang, E.~Oset and D.~M.~Li,
The $B^{+} \rightarrow J/\psi\omega K^{+}$ reaction and $D^{\ast} \bar{D}^{\ast}$ molecular states,
Eur. Phys. J. A \textbf{55} (2019) no.3, 36.
%doi:10.1140/epja/i2019-12706-6
%[arXiv:1808.10373 [hep-ph]].
%8 citations counted in INSPIRE as of 07 Feb 2021

%\cite{Romanets:2012hm}
\bibitem{Romanets:2012hm}
O.~Romanets, L.~Tolos, C.~Garcia-Recio, J.~Nieves, L.~L.~Salcedo and R.~G.~E.~Timmermans,
Charmed and strange baryon resonances with heavy-quark spin symmetry,
Phys. Rev. D \textbf{85} (2012), 114032.
%doi:10.1103/PhysRevD.85.114032
%[arXiv:1202.2239 [hep-ph]].
%97 citations counted in INSPIRE as of 14 Jan 2021

\end{thebibliography}

\end{document}